\documentclass[fleqn,11pt]{article}

\usepackage{amsfonts,amsmath,amssymb}
\usepackage{latexsym}
\usepackage{graphicx}
\usepackage{dcolumn}
\usepackage{bm}
\usepackage{hyperref}
\usepackage{enumitem}
\usepackage{accents}

\usepackage{amsfonts,amssymb,cite}
\usepackage{graphicx}

\def\noi{\noindent}
\def\jnumber#1#2{\thispagestyle{empty} \noi\unitlength=1mm
    	\begin{picture}(178,10)
            \put(177,15){\llap{\large\it Grav. Cosmol. No.\,#1, #2}}
                    \end{picture}}

\newcommand{\Title}[1]{\noi {{\Large\bf #1}}\\[1ex]}

\newcommand{\Author}[2]{\noi{\bf #1}\\[2ex]\noi{\normalsize\it #2}\\}

\def\au#1{${}^{#1}$}

\newcommand{\Abstract}[1]{\vskip 2mm \begin{center}
        \parbox{16.4cm}{\small\noi #1} \end{center}\medskip}

\def\email#1#2{\footnotetext[#1]{e-mail: #2}\addtocounter{footnote}{1}}

\def\nqq{\hspace*{-2em}}

\usepackage{color}

\def\Jl#1#2{#1 {\bf #2},\ }

\def\ApJ#1 {\Jl{Astroph. J.}{#1}}
\def\CQG#1 {\Jl{Class. Quantum Grav.}{#1}}
\def\DAN#1 {\Jl{Dokl. AN SSSR}{#1}}
\def\GC#1 {\Jl{Grav. Cosmol.}{#1}}
\def\GRG#1 {\Jl{Gen. Rel. Grav.}{#1}}
\def\IJMPD#1 {\Jl{Int. J. Mod. Phys. D}{#1}}
\def\JETF#1 {\Jl{Zh. Eksp. Teor. Fiz.}{#1}}
\def\JETP#1 {\Jl{Sov. Phys. JETP}{#1}}
\def\JHEP#1 {\Jl{JHEP}{#1}}
\def\JMP#1 {\Jl{J. Math. Phys.}{#1}}
\def\NPB#1 {\Jl{Nucl. Phys. B}{#1}}
\def\NP#1 {\Jl{Nucl. Phys.}{#1}}
\def\PLA#1 {\Jl{Phys. Lett. A}{#1}}
\def\PLB#1 {\Jl{Phys. Lett. B}{#1}}
\def\PRD#1 {\Jl{Phys. Rev. D}{#1}}
\def\PRL#1 {\Jl{Phys. Rev. Lett.}{#1}}

\def\lal{&&\nqq {}}

\def\beq{\begin{equation}}
\def\eeq{\end{equation}}
\def\bear{\begin{eqnarray}}
\def\bearr{\begin{eqnarray} \lal}
\def\ear{\end{eqnarray}}
\def\earn{\nonumber \end{eqnarray}}

\begin{document}
\twocolumn[
\jnumber{issue}{year}

\Title{A note on pure connection formalism for unimodular gravity and its possible generalisations}

\Author{Alexey L. Smirnov\au{1}}
       {Institute for Nuclear Research of the Russian Academy of Sciences,\\
         60-th October Anniversary Prospect, 7a, 117312, Moscow, Russia }


\Abstract{
  In this note, we consider  the Henneaux-Teitelboim version of Unimodular Gravity (UG) and its deformations
  in the form of gauge theories with spontaneously broken diffeomorphism invariance.
  Actions defining such theories depends on the curvature
  of an $SO(3,\mathbb{C})$ gauge connection and the field strength of a (real) 3-form (or equivalently its dual vector density).
  We obtain the pure connection action of the theory  from the corresponding Plebanski action by integrating
  out auxiliary fields. Then we show that the Henneaux-Teitelboim form of UG can be included in a wider class of theories
  which propagate
  two (complex) degrees of freedom.
}
\medskip
]
\email 1 {smirnov@inr.ac.ru}
\section{\label{sec:intro}Introduction}
It is known that General Relativity (GR) can be formulated as a diffeomorphism invariant gauge theory
of the $SO(3,\mathbb{C})$ connection without the explicit use of the metric~\cite{CDJ1,CDJ2,CDJ3,Krasnov1}.
In this approach the metric becomes a derived quantity which can be expressed in terms of the connection.
Moreover, such pure connection formulation allows to build the whole new class of gravitational theories
with two propagating degrees of freedom (DoF)~\cite{Krasnov2} and GR with non-zero cosmological constant arises
as a particular member of this class.

It appears that the pure-connection formulation for the Henneaux-Teitelboim
version~\cite{HT} of Unimodular Gravity does also exist~\cite{Smirnov,GN}.
Recall, that the term Unimodular Gravity refers in fact to a class of theories.
These theories are obtained from GR by imposing some specific constraints on the metric.
In the original form of UG, one set the metric determinant to a constant, thereby violating the full diffeomorphism group
to the subgroup of volume preserving (or unimodular) diffeomorphisms, hence the name of the theory.
The distinctive feature of UG theories is that the spontaneously broken diffeomorphism invariance leads to emergence
of a global DoF which show itself as a cosmological constant.

The Henneaux-Teitelboim form of UG  can be written as GR coupled with a 3-form gauge field.
All but one components of the 3-from are pure gauge. They serve as auxiliary fields which make the action of the theory 
manifestly diffeomorphism invariant. The remained mode of the 3-form is then a 'unimodular time' and its canonically conjugate
momentum is the cosmological constant. The unimodular time defines (on-shell) a preferred notion of time
thereby spontaneously violating diffeomorphism invariance.  
Nevertheless, the equations of motion lead to the evolution classically indistinguishable
form that of GR with cosmological constant.

As it was already mentioned, the action for Henneaux-Teitelboim form of UG (hereafter referred to as UG)
in the pure connection formalism has been recently obtained in~\cite{GN}.
In the Section~\ref{sec:pureconn}, we give an alternative derivation of this action presented
earlier in~\cite{Smirnov}. Our approach uses the UG action given by Smolin~\cite{Smolin1} as the starting point.
The Smolin's variant of the Plebansky action is described in the Section~\ref{sec:plebanski}.
In Section~\ref{sec:neighbours}, we describe a new class of deformed UG theories.
The Lagrangians of such theories are given by arbitrary homogeneous defining functions of degree one. Arguments of the defining
function are the field strength of the 3-form and the matrix defined as the wedge product of two curvature 2-forms
of the  $SO(3,\mathbb{C})$ connection. Our main goal is to show that such theories also contain one global and
two propagating (complex) DoF. To this end, we use the canonical formalism and construct the corresponding family
of Hamiltonians.   

It should also be stressed from the outset that since we work with the chiral (self-dual) formulation of Lorentzian UG,
quantities such as $SO(3,\mathbb{C})$ connection and curvature are inherently complex.
It means that we must supplement the actions discussed in the paper by some kind of reality conditions
in order to obtain real geometrodynamical quantities. The problem of reality conditions is in fact the main stumbling block
of the formalism. However, in the present paper, we do not touch this issue and simply assume
that such conditions do exist and can be consistently implemented. For recent discussion of the reality conditions
in  chiral diffeomorphism invariant gauge theories, see~\cite{Krasnov4}.

\section{\label{sec:plebanski}The Plebanski action for UG}
In this preparatory section we dsecribe the Smolin's formulation of unimodular gravity~\cite{Smolin1}
which is given by a modified Plebansky action. It uses slightly different set of auxiliary fields comparing
to the similar action of~\cite{GN}.

Let $M$ be an orientable four-dimensional manifold and we consider the following modified Plebansky action
\begin{equation}
  \begin{aligned}
    S_{\rm P}&=\frac{1}{8\pi G}\int_{M}\bigg[ F^i \wedge B_i\\
      &-\frac{1}{2}\left(\Psi_{ij}+\frac{\phi}{3}\delta_{ij}\right)B^i\wedge B^j+\rho\,\mbox{Tr}\Psi-2{\rm i\,}\phi\,b\bigg],
  \end{aligned}
  \label{BF}
\end{equation}
here $F^i=dA^i+(1/2)\varepsilon^{ijk}A^j\wedge A^k$ is the curvature of the~$SO(3,\mathbb{C})$ 1-form gauge connection $A^i$,
$B^i$ are three $\mathfrak{so}(3,\mathbb{C})$-valued 2-forms, $\Psi$ is the 3$\times$3 complex symmetric matrix.
Note, 'internal' group indices $i,j,\dots$=1,2,3 are raised and lowered by $\delta_{ij}$, therefore
their positions (upper or lower) are irrelevant. The Levi-Civita symbol $\varepsilon^{ijk}$ is the structure constant
of  $\mathfrak{so}(3,\mathbb{C})$ and \mbox{$\varepsilon^{123}=+1$}.
The quantity \mbox{$\rho=\tilde\rho\,d^4x$} where $\tilde\rho$ is a scalar density of weight one~\footnote{In the paper,
tensor densities of arbitrary positive weight are denoted by an appropriate number of tildes over the symbol.
For negative weights we use undertildes.} and $\phi$ is a scalar and they are two Lagrange multipliers.
The quantity $b=\tilde b\,d^4x$ where $\tilde b$ is the density of weight one dual to the field strength $b=da$ of the 3-form~$a$.
Thus the action~\eqref{BF} is invariant the under the additional gauge transformation
\begin{equation}
  a\to a+d\omega
  \label{agauge}
\end{equation} 
where $\omega$ is an arbitrary smooth 2-form.

Let us introduce a completely anti-symmetric rank four tensor density $\tilde\epsilon^{\mu\nu\lambda\rho}$ of weight one which
exists on any orientable manifold and does not require a metric for its definition. Due to the lack of the metric
in the formalism, $\tilde\epsilon^{\mu\nu\lambda\rho}$ and its covariant counterpart $\undertilde{\epsilon}_{\mu\nu\lambda\rho}$
have be used to raise and lower indices. Observe now, that there exists a unique anti-symmetric rank four tensor density
with the property that its components are constants in any coordinate system. This tensor density
is the Levi-Civita symbol and it is natural to choose it as $\tilde\epsilon^{\mu\nu\lambda\rho}$.

Then the convention $dx^{\mu}\wedge dx^{\nu}\wedge dx^{\lambda}\wedge dx^{\rho}=\tilde\epsilon^{\mu\nu\lambda\rho}dx^4$
allows us to write
\begin{equation}
  \begin{aligned}
    \tilde b&=\frac{1}{4!}\tilde\epsilon^{\mu\nu\lambda\rho}b_{\mu\nu\lambda\rho}d^4x=\partial_{\mu}\left(\frac{1}{3!}\tilde\epsilon^{\mu\nu\lambda\rho}a_{\nu\lambda\rho}\right)d^4x\\
    &=\partial_{\mu}\tilde a^{\mu}d^4x
      \end{aligned}
\end{equation}
where $\tilde a^{\mu}$ is the vector density dual to the 3-form~$a$.

Variation with respect to~$\tilde a^{\mu}$ shows that $\phi$ is a constant on shell
\begin{equation}
  \phi(x)=\Lambda.
  \label{phieq}
\end{equation}
Varying $\Psi_{ij}$ the 'metricity' equation can be found
\begin{equation}
  B^i\wedge B^j-\frac{1}{3}\delta_{ij}B_i\wedge B^i=0.
  \label{meteqBF}
\end{equation}
Solutions to this equations are linear combinations of simple bivectors composed of
the tetrad components $\theta^0$, $\theta^i$
\begin{equation}
  B^i=\left({\rm i}\theta^0\wedge\theta^i-\frac{1}{2}\epsilon^{ijk}\theta^j\wedge\theta^k\right)
  \label{B2tetrad}
\end{equation}

The unimodular condition follows if we vary the action~\eqref{BF} with respect to the field $\phi$
\begin{equation}
  b = \frac{\rm i}{6}B_i\wedge B^i.
  \label{umc}
\end{equation}
It is now clear the meaning of the prefactor ${\rm i}$ in the last term of~\eqref{BF}; it makes the volume form real.

It follows from \eqref{B2tetrad} and \eqref{umc} that \mbox{$\sqrt{-g}=\tilde b$} on-shell. Observe, because the form $b$
is exact, then  by the Sokes' theorem the volume of any bounded spacetime region is given by integral of 3-form $a$
over the boundary. For example, let us for a moment assume that $M$ is a spatially compact manifold and we selected
two nonintersecting spacelike hypersurfaces $\Sigma_1$, $\Sigma_2$. Integration of \eqref{umc} over the spacetime
region ${\cal R}$ bounded by these hypersurfaces gives
\begin{equation}
  \mbox{vol}({\cal R})=\tau(\Sigma_2)-\tau(\Sigma_1),\quad \tau(\Sigma):=\int_{\Sigma}\,a
\end{equation}
where  $\tau(\Sigma)$ is the elapsed spacetime volume to the past of $\Sigma$. Now, $\tau(\Sigma)$ can be considered
as a global time coordinate that breaks the diffeomorphism invariance on-shell. 

Further, variation of~\eqref{BF} with respect to $A^i$ gives
\begin{equation}
  DB^i=dB^i+\epsilon^{ijk}A^j\,B^k=0
\end{equation}
Solving this equation algebraically for $A^i$ and implying that $B^i$ is a solution of~\eqref{meteqBF} we recover the self-dual
part of the torsion-free spin connection. 

At last, by varying $B^i$ and taking into account~\eqref{phieq} we can write the Einstein equations in the form
\begin{equation}
  F_i=\left(\Psi_{ij}+\frac{\Lambda}{3}\delta_{ij}\right)B^i
\end{equation}
Thus, on the equation of motion, the matrix $\Psi$ becomes the self-dual part of the Weyl curvature.

\section{\label{sec:pureconn} Pure connection formulation of UG}
Now taking the action~\eqref{BF} as stating point, we can adopt the procedure described in~\cite{CGM} in order to
obtain the action principle for the pure connection formulations of UG.

First of  all, let us make the following redefinition
\begin{equation}
\Phi_{ij}=\Psi_{ij}+\frac{\phi}{3}\delta_{ij}.
\end{equation} 
Then varying~\eqref{BF} with respect to $B^i$ and {\it assuming} that $\Phi_{ij}$ is invertible, we can write
$\mbox{$B_i=\Phi^{-1}_{ij}F^j$}$ and then substitute it back to~\eqref{BF} to obtain
\begin{equation}
  \begin{aligned}
    S=\frac{1}{8\pi G}\int_{M}\bigg[\frac{1}{2}\Phi^{-1}_{ij}\tilde X^{ij} &+\frac{\tilde\rho\,\mbox{Tr}\Phi}{2}\\
      &-\frac{\phi\,\tilde\rho}{2}-2{\rm i}\,\phi\,\tilde b \bigg]d^4x,
    \end{aligned}
  \label{preUG}
\end{equation} 
where the symmetric $\tilde X^{ij}$ has been defined according to
\begin{equation}
F^i\wedge F^j=\tilde X^{ij}\,d^4x=\frac{1}{4}\tilde\epsilon^{\mu\nu\lambda\rho}F^i_{\mu\nu}F^j_{\lambda\rho}d^4x.
\end{equation}
Now, varying $\Phi$ in the above action gives rise to the matrix equation
\begin{equation}
  \tilde X-\tilde\rho\,\Phi^2=0
  \label{PhiXeq}
\end{equation} 
Since by assumption $\Phi$ is invertible, then \eqref{PhiXeq} implies invertibility of $\tilde X$.
Therefore we can solve \eqref{PhiXeq} with respect to $\Phi$ by taking a square root of $X$
\begin{equation}
  \Phi=\tilde\rho^{-1/2}\tilde X^{1/2}
  \label{PhiXsol}
\end{equation}
It should be stressed that the solution \eqref{PhiXsol} is not unique since the square root of $\tilde X$ is not unique.
However the procedure of obtaining new action is not affected by this ambiguity.

Inserting \eqref{PhiXsol} into \eqref{preUG} and the eliminating Lagrange multipliers $\tilde\rho$ and $\phi$ by means of their
equations of motion we eventually arrive to the main result of the section
\begin{equation}
  S_{\rm PC}=\frac{\sqrt{{\rm i}}}{4\pi G}\int_{M} d^4x \sqrt{\tilde b}\,\mbox{Tr}\sqrt{\tilde X}
  \label{PCUG}
\end{equation}
The integrand in \eqref{PCUG} is a scalar density of weight one and the action is invariant under diffeomorphisms.
In principle, the action \eqref{PCUG} coincides with one of the actions constructed in~\cite{GN} (cf. Eq.22).
However, the authors of~\cite{GN} fix the volume form by means of unimodlular condition.

Varying of \eqref{PCUG} with respect to $A^i$ and  $\tilde a^{\mu}$ gives the equations of motion
\begin{subequations}
  \begin{flalign}
    D\left(\sqrt{\tilde b}\,(\tilde X^{-1/2})^{ij}F^j\right)&=0,\label{Feq1}\\
    \partial_{\mu}\left(\frac{\mbox{Tr}\sqrt{\tilde X}}{\sqrt{\tilde b}}\right)&=0\label{beq1}
    \end{flalign}
\label{eoms1}
\end{subequations}
Observe, that in the first equation of~\eqref{eoms1} we have used the expression
\begin{equation}
  \frac{\partial\mbox{Tr}\sqrt{\tilde X}}{\partial \tilde{X}^{ij}}=\frac{1}{2}(\tilde{X}^{-1/2})^{ij}
\end{equation}

Let us shortly discuss properties of the theory which follow from~\eqref{eoms1}.
First of all, integrating \eqref{beq1} we introduce a {\it complex} integration constant $\Lambda$ into the theory.
When $\Lambda\neq 0$ we can substitute the integrated equation into \eqref{Feq1} and formally reproduce the equation of motion
in the pure connection formulation of GR \cite{Krasnov1} but now with complex cosmological constant.
Given the fact that $\tilde b$ must be real, the integrated \eqref{beq1} leads to the following reality condition
\begin{equation}
  \mbox{Im}\left(\frac{\mbox{Tr}\sqrt{\tilde X}}{\Lambda}\right)=0
  \label{RC}
\end{equation} 

In the special case when $\Lambda=0$, the second equation gives
\begin{equation}
\mbox{Tr}\sqrt{\tilde X}=0,
\end{equation}
provided that $\tilde b$ is non-singular. In fact, $\tilde b$ can now be considered as a Lagrange multiplier and
the unimodular condition is lost.
The theory can be reduced to the pure connection formulation of GR described in~\cite{CDJ1,CDJ2}.

Having know $F^i$, the metric can be reconstructed in the form of the Urbantke metric~\cite{Urbantke}
\begin{equation}
  g_{\mu\nu}=\undertilde{c}_{F}\,\epsilon_{ijk}\tilde\epsilon^{\alpha\beta\gamma\delta}F^i_{\mu\alpha}F^j_{\beta\gamma}F^k_{\delta\nu},
  \label{Umetric}
\end{equation}
where the conformal factor $\undertilde{c}_F$ is a density of weight~-1. Note, that the similar metric
can also be constructed in the Plebasky formalism by using the forms $B^i$. In this case $\undertilde{c}_B=1/\tilde{b}$
because of~\eqref{umc}. Since \mbox{$B^i\sim \sqrt{\tilde b}\,(\tilde{X}^{-1/2})^{ij}\,F^j$},
we can compare two metrics to obtain $\undertilde{c}_F$ up to an overall constant factor
\begin{equation}
  \undertilde{c}_F=\sqrt{\frac{\tilde b}{\mbox{det}\tilde X}}.
\end{equation} 

In this way we have got access to geometrodynamical quantities. However, as it was stressed
in the Introduction, they are inherently complex and addtional reality conditions must be added to the theory.

\section{\label{sec:neighbours} Deformations of UG}
The theory described in the previous section allows for a generalisation. For example, it is easy to see that the action
\begin{equation}
  S^{(\alpha)}_{\rm PC}\sim\int_{M} d^4x\,\tilde{b}^{\alpha}\,(\mbox{Tr}\sqrt{\tilde X})^{2(1-\alpha)},\quad \alpha\in\mathbb{C}\backslash\{0,1\}
  \label{alphaPCUG}
\end{equation}
leads to {\it the same} dynamical EoMs as the theory of the previous section but
with the replacement $\Lambda\to\Lambda^{\frac{\alpha}{1-\alpha}}$. This observation suggests that there must exist theories
similar to UG which contain the same number of DoF. 

Therefore the next obvious step is to consider the general action
\begin{equation}
  S=\int_{M}\,f(b, F^i\wedge F^j)=\int_{M} d^4x\,f(\tilde b, \tilde X^{ij})
  \label{fPCUG}
\end{equation}
where the defining function $f$ has the properties: i) it is homogeneous function of degree one; ii) it is holomorphic
in the second argument and the matrix $\partial f/\partial \tilde X^{ij}$ is not singular;
iii) it is an $SO(3,\mathbb{C})$ invariant. 

The conditions (i) and (iii) implies that  the action \eqref{fPCUG} should be invariant with respect to diffeomorphism
and gauge transformations just by design. However, following~\cite{Krasnov2}, one can nevertheless perform explicit calculation
to show that these invariances hold. The diffeomorpism and gauge transformations act on the fields as follows
\begin{subequations}
  \begin{flalign}
    \delta_{\xi}A^i_{\mu}&=-\xi^{\nu}F^i_{\mu\nu},\\
    \delta_{\phi}A^i_{\mu}&=D_{\mu}\phi^i,\\
    \delta_{\xi}\tilde a^{\mu}&={\cal L}_{\xi}\tilde a^{\mu},\\
    \delta_{\omega}\tilde a^{\mu}&=\frac{1}{3!}\tilde \epsilon^{\mu\nu\lambda\rho}\partial_{[\nu}\omega_{\lambda\rho]}
  \end{flalign}
  \label{vars}
\end{subequations}
The first transformation is just the Lie derivative of $A^i_{\mu}$ along $\xi\in \mathfrak{diff}(M)$ modulo an
$SO(3,\mathbb{C})$ gauge transformation,
the second one is an $SO(3,\mathbb{C})$  rotation with the parameter $\phi^i$,
the third one is the Lie derivative of $\tilde a^{\mu}$ and the fourth one is the gauge transformation of $\tilde a^{\mu}$
with the parameter $\omega$, see \eqref{agauge}. Now, calculations of~\cite{Krasnov2} (see Sec 2.7. there) can be applied
almost literally to~\eqref{fPCUG} and the total variation of the action reads
\begin{equation}
  \delta S= \int_{M} d^4x\,\partial_{\mu}\left(f\,\xi^{\mu}\right).
  \label{varS}
\end{equation}
Here the terms corresponding to the gauge transformations vanish algebraically, and \eqref{varS} is just a boundary term,
i.e it has also correct behaviour under diffeomorphisms.

In order to count DoF, it is instructive to consider the canonical formalism for the theories~\eqref{fPCUG}.
To this end, we perform 3+1 decomposition of the action implying that the manifold is foliated by
closed\footnote{The closedness of~$\Sigma$ is merely a simplifying assumption. It allows us not deal with
boundary terms.} spatial hypersurfaces~$\Sigma$ i.e. $M=\mathbb{R}\times\Sigma$.
The lowercase letters from the beginning of the alphabet denote
spatial indices, while 0 denote temporal index.

Then 3+1 decomposition of the matrix $\tilde X$
\begin{equation}
  \tilde X^{ij}=\frac{1}{2}\tilde\epsilon^{abc}\left(F_{0a}^i F_{bc}^j+F_{0a}^j F_{bc}^i\right)
\end{equation} 
along with the Euler's theorem for homogeneous functions allows us to rewrite the action in the following form
\begin{equation}
   \begin{aligned}
     S=\int_{\mathbb{R}}dt\int_{\Sigma}d^3x\big(\tilde\Pi_i^a\dot A_a^i &+ \Pi\,\dot{\tilde a}^0\\
     &+\tilde a^b\,\partial_b\,\Pi+A_0^iD_a\tilde\Pi_i^a \big),
     \end{aligned}
  \label{fPCUG3p1}
\end{equation}
where $\tilde\Pi_i^a$, $\Pi$ are canonical momenta conjugate to $A_a^i$, $\tilde a^0$ respectively and defined as 
\begin{equation}
    \tilde\Pi_i^a:=\frac{\partial f}{\partial \tilde X^{ij}}F_{bc}^i\tilde\epsilon^{abc},\quad\Pi:=\frac{\partial f}{\partial\tilde b}\,,
  \label{momenta}
\end{equation}
and the last two terms in~\eqref{fPCUG3p1} are proportional to two Gauss constraints
\begin{equation}
  {\cal G}_a^{(a)}:=\partial_a\Pi,\quad {\cal \tilde G}_i^{(A)}:=D_a\tilde\Pi_i^a.
  \label{gaussC}
\end{equation}

The action~\eqref{fPCUG3p1} is still not complete since the first class constraints which generate the diffeomorphism
symmetry are absent.  In the theory under study the constraints generating diffeomorphisms of $\Sigma$ reads
\begin{equation}
  \tilde V_b\equiv\tilde\Pi_i^aF^i_{ab}=\frac{\partial f}{\partial \tilde X^{ij}}F_{cd}^i\,F_{ab}^j\tilde\epsilon^{acd}=0
  \label{diffC}
\end{equation} 
and it is in fact an algebraic constraint because it follows from the identity $F_{ab}^{(i}F_{cd}^{j)}\tilde\epsilon^{acd}=0$.
Note, the constraint~\eqref{diffC} is the same as for GR written in the Ashtekar variables~\cite{Ashtekar1}.

As a preliminary step in obtaining the Hamiltonian constraint observe that from the expression
$3!\,\mbox{det}\,\tilde\Pi=\epsilon^{ijk}\undertilde\epsilon_{abc}\tilde\Pi_i^a\,\tilde\Pi_j^b\,\tilde\Pi_k^c$, and \eqref{momenta}
we can deduce
\begin{equation}
  \Psi_{ij}^{-1}=\frac{\partial f}{\partial \tilde X^{ij}},\quad\Psi_{ij}:=\frac{\epsilon^{ikl}\tilde\Pi_k^a\,\tilde\Pi_l^b\,F_{ab}^j}{\mbox{det}\,\tilde\Pi}
  \label{invPsi}
\end{equation}
Then it seems that the only possible form of the Hamiltonian constraint which properly closes the hypersurface deformation
algebra must have the form
\begin{equation}
  \tilde{\tilde H}=\epsilon^{ijk}\tilde\Pi_i^a\,\tilde\Pi_j^b\,F_{ab}^k+{\cal H}(\Pi,\Psi_{\rm TF})\,\mbox{det}\,\tilde\Pi\approx 0
  \label{hamC}
\end{equation}
Here $\Psi_{\rm TF}$ is the trace-free part of~$\Psi$ and the symbol '$\approx$' means that the expression vanishes on
the constraint surface. It follows from~\eqref{invPsi} and~\eqref{hamC} that
\begin{equation}
  {\cal H}(\Pi,\Psi_{\rm TF})+\mbox{Tr}\,\Psi\approx 0,
  \label{H2f}
\end{equation}
which relates the function~$\cal H$ and the defining function~$f$. Thus knowing $f$ we can reconstruct $\cal H$.
For example, the theories~\eqref{alphaPCUG} are recovered when ${\cal H}\sim\Pi^{\frac{\alpha}{1-\alpha}}$.
Similarly to the case of GR~\cite{Krasnov3}, the class of theories with Hamiltonians~\eqref{hamC}
can be dubbed as 'deformations' of UG.

If we define the smeared Hamiltonian constraint as the functional
\begin{equation}
  {\cal C}_{\undertilde N}:=\int_{\Sigma} dx^3 \undertilde N  \tilde{\tilde H}
  \label{hamCfun}
\end{equation}
where $\undertilde N$ is the densitizied lapse function, then the Poisson brackets reads as
\begin{equation}
  \begin{aligned}
    \{{\cal C}_{\undertilde{N}_1},{\cal C}_{\undertilde{N}_2}\}&:=\int_{\Sigma} d^3x\bigg(\frac{\delta {\cal C}_{\undertilde{N}_1}}{\delta  A_a^i}\frac{\delta {\cal C}_{\undertilde{N}_2}}{\delta \tilde\Pi_i^a}-\frac{\delta {\cal C}_{\undertilde{N}_2}}{\delta  A_a^i}\frac{\delta {\cal C}_{\undertilde{N}_1}}{\delta \tilde\Pi_i^a}\\
    &+\frac{\delta {\cal C}_{\undertilde{N}_1}}{\delta \tilde{a}^0}\frac{\delta {\cal C}_{\undertilde{N}_2}}{\delta\Pi}-\frac{\delta {\cal C}_{\undertilde{N}_2}}{\delta \tilde{a}^0}\frac{\delta {\cal C}_{\undertilde{N}_1}}{\delta\Pi}\bigg).
  \end{aligned}
\end{equation}
The diffeomorphism invariance dictates that the Poisson brackets of ${\cal C}_{\undertilde N}$ with itself must be proportional
to the smeared constraints~\eqref{diffC}.
To proof this, the procedure described in~\cite{Krasnov3} can be applied to~\eqref{hamCfun} without modifications
because~\eqref{hamC} does not depend on the canonical coordinate~$\tilde{a}^0$ and therefore ($\tilde{a}^0$, $\Pi$) part of the Poisson brackets does not contribute. Then we have 
\begin{equation}
  \begin{aligned}
\{{\cal C}_{\undertilde{N}_1},{\cal C}_{\undertilde{N}_2}\}&=4\int_{\Sigma}dx^3\,\tilde{\tilde q}^{ab}\undertilde{\undertilde N}_b\,\tilde V_a,\\
    \undertilde{\undertilde N}_a&:=\partial_a\undertilde{N}_1\,\undertilde{N}_2-\partial_a\undertilde{N}_2\,\undertilde{N}_1,\\
    \tilde{\tilde q}^{ab}&:=\frac{1}{2}\epsilon^{ijk}\epsilon^{lmn}\tilde\Pi_i^a\,\tilde\Pi_l^b\,h_{jm}\,h_{kn},\\
    h_{ij}&:=\delta_{ij}+\frac{\partial{\cal H}(\Pi, \Psi_{\rm TF})}{\partial \Psi_{\rm TF}^{ij}},
  \end{aligned}
  \label{pb4C}
\end{equation}
We note that in the Ashtekar formalism, $\tilde{\tilde q}^{ab}$ defines the (densitized)
metric on the hypersurface $\Sigma$ and momenta $\tilde\Pi_i^a$ define a (densitizied) triad.
However, the metric  $\tilde{\tilde q}^{ab}$ in~\eqref{pb4C} is different from that of GR (and the UG of the previous section)
where it is simply $\tilde\Pi_i^a\,\tilde\Pi_i^b$. Instead, $\tilde\Pi_i^a$'s are now contracted
with the non-flat 'internal' metric~$h_{ij}$.

The result~\eqref{pb4C} is sufficient to proof that that the algebra of constraints is closed because 
the Poisson brackets of ${\cal C}_{\undertilde N}$ with smeared  ${\cal \tilde G}_i^{(A)}$ and  $\tilde V_a$
remains the same as in the case of GR.
Indeed, the Hamiltonian is gauge invariant and therefore it commutes with the Gauss constraints,
while the Poisson brackets of smeared Hamiltonian and diffeomorphism constraints return
the Hamiltonian constraint smeared with the Lie derivative of the lapse function
along the shift vector~\cite{Ashtekar1,Krasnov3}. 

Having defined ${\cal G}_a^{(a)}$,  ${\cal \tilde G}_i^{(A)}$, $\tilde V_a$, $\tilde{\tilde H}$, we obtain
the algebra of the first class constraint of the theory and the complete 3+1 decomposed action can be written as follows
\begin{equation}
  \begin{aligned}
    S&=\int_{\mathbb{R}}dt\int_{\Sigma}d^3x\big(\tilde\Pi_i^a\dot A_a^i+\Pi\,\dot{\widetilde a}^0\\
    &+\widetilde a^b\,{\cal G}_b^{(a)}+A_0^i{\cal \tilde G}_i^{(A)}+N_a\tilde V^a+\undertilde N \tilde{\tilde H} \big)
  \end{aligned}
  \label{PCUG3p1}
\end{equation}  

We can now count DoF of the theory. We have 1+3$\times$3=10 configuration variables ~$\tilde{a}^0$, $A_a^i$
and the set of 1+3+3+1=8 first class constraints. Thus we obtain 10-8=2 (complex) DoF per a spacetime point.
As expected, the 3-form field $a$ does not have propagating degrees of freedom but rather provides the existence
of a constant of the motion~$(\dot\Pi,\partial_a\Pi)=\partial_{\mu}\Pi=0$ as it follows from~\eqref{PCUG3p1}.

\section{\label{sec:concl} Conclusions and outlook}
In this paper we shortly discussed some basic properties of UG and its generalisations in the pure connection formulation.
First of all, we gave a derivation of the action for UG alternative to~\cite{GN}.
More importantly, we presented a wider class of deformed UG theories~\eqref{fPCUG}
which also contains two propagating (complex) DoF. 

Note that the theories~\eqref{fPCUG} may lead to the same classical dynamics
as those discussed in~\cite{Krasnov2}. Therefore, as the prospect for the future work,
it would be interesting to study the perturbative quantum dynamics of such theories.
We expect that the approach applied in~\cite{Krasnov5} to the deformations of GR can presumably be adopted to~\eqref{fPCUG}.

The issue of the reality conditions is yet another important direction of the future research.
We can roughly formulate this issue as follows.
We can use the Urbantke metric to construct the geometrodynamical description of the theories~\eqref{fPCUG}.
If such description do exist then the metric on the spacelike slices together with lapse and shift functions must be real
during the time evolution. These conditions must then be translated into some reality conditions for canonical variables.
At present, it is not clear at all if the consistent set of the reality conditions do exist and lead (along with the constraints)
to two real DoF in the theories of type~\eqref{fPCUG}.

Finally, we note that there may exist relation between the pure connection formulation of UG and recently proposed  generalized UG~\cite{Barvinsky1,Barvinsky2,Nesterov1}. This putative relation also deserves a separate study. In the metric formalism the generalized UG follows from the Einstein-Hilbert action with the additional constraint that the lapse~$N$ becomes an arbitrary function of determinant of the spatial metric. This constraint breaks the diffeomorphism invariance on-shell and such theories look like GR plus a perfect fluid. At the moment we can point out two possibilities which may provide a connection between the present work and the  generalized UG. For example, one can somehow to extend~\eqref{fPCUG} and obtain a  pure connection version of the generalized UG. Another possibility is again related with reality conditions. It may be that the consistent set of reality conditions imposed on the theory~\eqref{fPCUG} itself would lead to a version of  the generalized UG. This latter approach can be justified by the following arguments. It is known that deformations of GR similar to~\eqref{fPCUG} can be rewritten in the form of complex GR plus a set of (complex) scalar fields~\cite{Krasnov4}. We expect that similar representation can be constructed for the theory~\eqref{fPCUG} itself. Then by imposing the reality conditions we could presumably mix complex DoF in such a way that the real section of~\eqref{fPCUG} would lead to the generalized UG of~\cite{Barvinsky1,Barvinsky2,Nesterov1}.

\end{document}